# TOP MOMENTUM RECONSTRUCTION AND EXTRA SOFT JETS[*]

LYNNE H. ORR [†]
*Department of Physics and Astronomy, University of Rochester,
Rochester, NY 14627, USA*

and

W.J. STIRLING
*Departments of Physics and Mathematical Sciences, University of Durham
Durham DH1 3LE, England*

## ABSTRACT

Top pairs produced at the Tevatron may be accompanied by extra jets due to radiation of soft gluons. Whether these jets should be included in top momentum reconstruction depends on the source of the gluons, which can be radiated off the initial partons, the final state $b$ quarks, or the intermediate top quarks. We compute soft gluon distributions and discuss their implications for top mass measurements.

Top quark pairs produced at the Tevatron may be accompanied by additional soft jets due to radiation of soft gluons. In attempts to measure the top mass $m_t$ by reconstructing the top quark's four–momentum from the momenta of its decay products, the question of how to deal with additional soft jets must be addressed. In particular, should such jets be combined with the $W$ and $b$ in reconstructing $m_t$? Obviously, if the gluon was radiated off an initial state quark, the answer is "no," and if it was radiated off one of the final $b$'s, the answer is "yes." Correspondingly, we would guess that soft jets near the beam come from initial state radiation, and thus should be ignored, and that central jets are due to final state radiation from $b$ quarks, and should be included. But what about gluons radiated off the top quarks themselves? Although top decays before hadronizing, it does have time to radiate gluons. Do such gluons belong to the initial or final state?

Clearly, an initial/final state interpretation is too naïve, and we must consider top production and decay simultaneously in our treatment of soft gluon radiation; *i.e.*, we must include all possible diagrams. Fortunately, the result can be decomposed into contributions associated with $t\bar{t}$ production, $t$ and $\bar{t}$ decay, and their interference. (See Ref. [1] for a discussion of the associated formalism.) This is exactly the decomposition we need for purposes of momentum reconstruction.

---

[*]Presented by L.H. Orr at DPF'94, Albuquerque, NM, August 2–6, 1994.
[†]Work supported in part by the U.S. Department of Energy, grant DE-FG02-91ER40685.

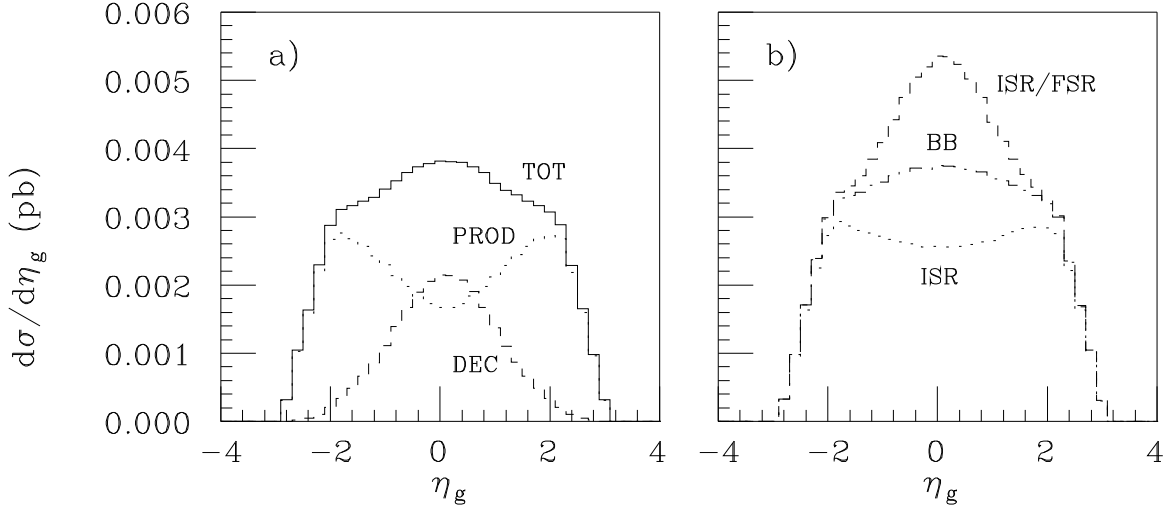

Fig. 1. Gluon pseudorapidity distributions in $t\bar{t}$ production via $q\bar{q} \to t\bar{t} \to b\bar{b}W^+W^-$, in $p\bar{p}$ collisions at $\sqrt{s} = 1.8$ TeV. (a) Net distribution and contributions from production and decay. (b) Distributions arising from ISR, ISR/FSR, and BB models described in the text.

Consider the process $q\bar{q} \to t\bar{t} \to bW^+\bar{b}W^-$ with emission of a gluon. In the soft approximation, the lowest order cross section and gluon distribution factorize, and we have $(1/d\sigma_0)\, d\sigma/(dE_g\, d\cos\theta_g\, d\phi_g) = \frac{\alpha_s}{4\pi^2} E_g\, (\mathcal{F}_{\rm PROD} + \mathcal{F}_{\rm DEC} + \mathcal{F}_{\rm INT})$. $\mathcal{F}_{\rm PROD}$ corresponds to gluons radiated in association with $t\bar{t}$ production, i.e., radiated before the $t$ or $\bar{t}$ quark goes on shell. Similarly, $\mathcal{F}_{\rm DEC}$ corrseponds to gluons radiated in the decay of the $t$ or $\bar{t}$. $\mathcal{F}_{\rm INT}$ represents the interferences between the two and depends on the top width $\Gamma_t$. Expressions for $\mathcal{F}_{\rm PROD}$, $\mathcal{F}_{\rm DEC}$, and $\mathcal{F}_{\rm INT}$ can be found in Ref. [2]. This production–decay–interference decomposition determines for us whether the gluon momentum should be combined with those of the $W$ and $b$ to form the top momentum. Gluons from the production piece are not to be included, while gluons from the decay piece are. For the interference term, whether to add the gluon is undetermined, but this contribution is negligible in the case of interest here.

Let us now obtain the full soft gluon distribution for $t\bar{t}$ production at the Tevatron. The results shown here are from Ref. [3], to which the reader is referred for a more complete discussion. We consider angular distributions since we want to know where the soft jets are expected to appear in the detector. We use $m_t = 174$ GeV, work at the parton level, and keep kinematic cuts to a minimum. The cuts are $|\eta_b|, |\eta_{\bar{b}}| \leq 1.5$; $|\eta_g| \leq 3.5$; $10$ GeV/$c \leq p_T^g \leq 25$ GeV/$c$; $E_g \leq 100$ GeV; and $\Delta R_{bg}, \Delta R_{\bar{b}g} \geq 0.5$. The resulting distribution in the gluon pseudorapidity is shown in Fig. 1(a) as a solid line, along with the contributions from production (dotted line) and decay (dashed line). We see that, as we might expect, the decay piece is strongly peaked in the center, and the production piece contributes most of the gluons in the forward and backward directions. Note, however, that gluons in the central region are almost as likely to have come from production as from decay, hence we cannot simply assign central soft jets to the decay piece.

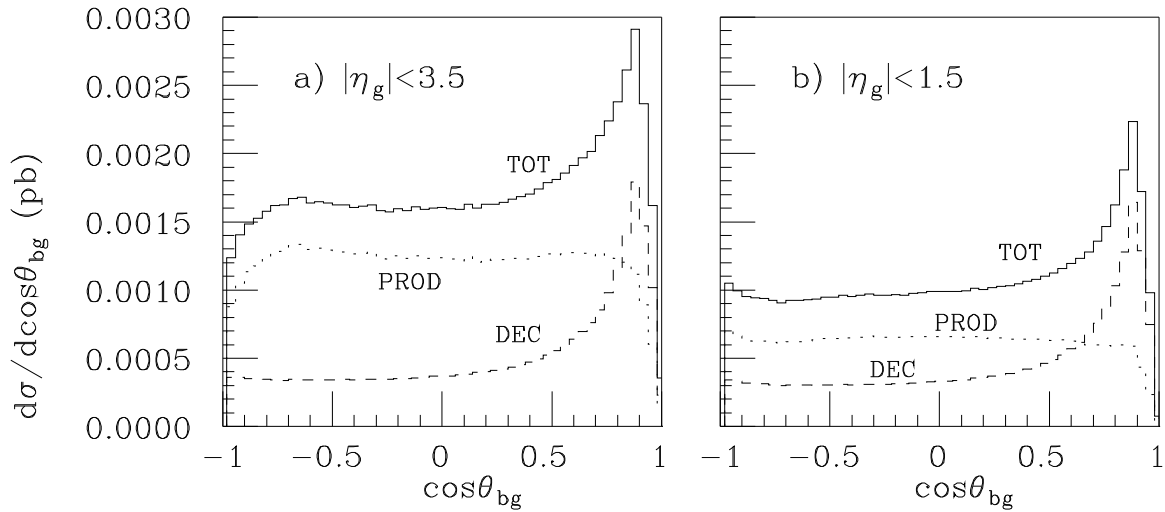

Fig. 2. Distribution in the cosine of the angle between the gluon and the $b$-quark, (a) with cuts described in the text and (b) with the additional cut $|\eta_g| \leq 1.5$.

We can hope to get a better handle on how to assign soft jets by using proximity to the $b$ quark. Figure 2(a) shows the distribution in cosine of the angle between the gluon and the $b$ quark with the same decomposition as in Fig. 1(a). We see a slight enhancement near the $b$ in the decay piece compared to production, but the difference is not terribly dramatic and is very sensitive to the cut on $\Delta R_{bg}$. In addition, it must be kept in mind that no fragmentation or detector effects have been included. The situation improves a little if we consider only central gluons, as shown in Fig. 2(b), where we take $|\eta_g| \leq 1.5$. The excess in the decay piece is now more pronounced.

Finally, we compare the correct pseudorapidity distribution in Fig. 1(a) to those in Fig. 1(b), obtained from some simpler models that are intuitively appealing and easily implemented in Monte Carlo simulations. The ISR model (dotted line) includes radiation off the initial $q\bar{q}$ state only, as if the $q$ and $\bar{q}$ formed a color singlet. We might expect this to correspond to the contribution associated with production, but we see by comparing to the dotted line in Fig. 1(a) that the ISR model overestimates radiation in the central region. In the ISR/FSR model (dashed line) we add to the ISR model radiation from the final $b\bar{b}$ pair as if they too formed a color singlet. This model corresponds roughly to the naïve expectation mentioned in the introductory paragraph. Figure 1(b) shows that this model overestimates the total radiation and gets the shape wrong. In the BB model (dot-dashed line) we use the correct color structure but ignore radiation off the top quark. This model approximately reproduces the correct pseudorapidity distribution. However, it does not give the correct azimuthal distribution,[3] and, more important for $m_t$ reconstruction, does not permit a production–decay decomposition.

In conclusion, gluon radiation in $t\bar{t}$ production and decay has a structure which gives rise to interesting and sometimes subtle effects, a few of which we have illustrated here. We have seen that there is no simple prescription for treating extra soft jets in $m_t$ reconstruction, but the production–decay decomposition provides some guidance. We have also shown that simpler, intuitively appealing models either do not reproduce the correct distributions, or do not allow for the production–decay decomposition.